\documentstyle[preprint,aps,graphicx]{revtex}

\begin{document}
\draft

\tightenlines
\title{Josephson Junction Arrays with Bose-Einstein Condensates}

\author{F. S. Cataliotti$^{1}$,
S. Burger, 
C. Fort, 
P. Maddaloni$^{2}$,\\
F. Minardi, 
A. Trombettoni$^{3}$, 
A. Smerzi$^{3}$, 
M. Inguscio$^{1}$}
\address{
 INFM and European Laboratory for Non-Linear Spectroscopy (LENS)\\
L.go~E.~Fermi~2, I-50125 Firenze, Italy\\
1) also Dipartimento di Fisica, Universit\`a di Firenze,
L.go~E.~Fermi~2, I-50125 Firenze, Italy\\
2) also Dipartimento di Fisica, Universit\`a di Padova,
via~F.~Marzolo~8, I-35131 Padova, Italy\\
3) International School for Advanced Studies (SISSA),
via Beirut 2/4, I-34014 Trieste, Italy}
\maketitle
\bigskip
\bigskip
{We report on the direct observation of an oscillating atomic
  current in a one-dimensional array of Josephson junctions realized
  with an atomic Bose-Einstein condensate. The array is created by a
  laser standing-wave, with the condensates trapped in the valleys of
  the periodic potential and weakly coupled by the inter-well
  barriers.  The coherence of multiple tunneling between adjacent
  wells is continuously probed by atomic interference.  The square of
  the small-amplitude oscillation frequency is proportional to the
  microscopic tunneling rate of each condensate through the barriers,
  and provides a direct measurement of the Josephson critical current
  as a function of the intermediate barrier heights. Our superfluid
  array may allow investigation of phenomena so far inaccessible to
  superconducting Josephson junctions and lays a bridge between the
  condensate dynamics and the physics of discrete nonlinear media.}
\newpage

The existence of a Josephson current through a potential barrier
between two superconductors or between two superfluids is a direct
manifestation of macroscopic quantum phase coherence {\it (1,2)}.  The
first experimental evidence of a current-phase relation was observed
in superconducting systems soon after the Josephson effect was
proposed in 1962 {\it (3)}, whereas verification in superfluid Helium
has been presented only recently owing to the difficulty of creating
weak links in a neutral quantum liquid {\it (4,5)}.  The experimental
realization of Bose-Einstein condensates (BEC) of weakly interacting
alkali atoms {\it (6,7)} has provided a route to study neutral
superfluids in a controlled and tunable environment {\it (8,9)} and to
implement novel geometries for the connection of several Josephson
junctions so far unattainable in charged systems.  The possibility of
loading a BEC in a one-dimensional periodic potential has allowed the
observation of quantum phase effects on a macroscopic scale such as
quantum interference {\it (10)} and the study of
superfluidity on a local scale {\it (11)}.
      
A Josephson junction (JJ) is a simple device made of two
coupled macroscopic quantum fluids {\it (2)}.  If the coupling is
weak enough, an atomic mass current $I$ flows across the two systems,
driven by their relative phase $\Delta \phi$ as:
\begin{equation}
\label{current-phase}
I=I_c \sin{\Delta \phi}
\end{equation}
where $I_c$ is the ``Josephson critical current'', namely  
the maximal current allowed to flow through the junction.
The relative phase dynamics, on the other hand, is sensitive to the 
external and internal forces driving the system:  
\begin{equation}
\label{ac}
\hbar \frac{d}{dt}\Delta \phi = \Delta V 
\end{equation}
where $\hbar$ is Planck's constant divided by $2 \pi$, $t$ is time,
and $\Delta V$ is the chemical potential difference between the
two quantum fluids. Arrays of JJs are made of several
simple junctions connected in various geometrical configurations. In
the past decade, such systems have attracted much interest, because of
their potential for studying quantum phase transitions in systems
where the external parameters can be readily tuned {\it (12)}.
Recently, the creation of simple quantum-logic units and more complex
quantum computer schemes {\it (13)} have been discussed.  A great
level of accuracy has been reached in the realization of two- and
three-dimensional superconducting JJ arrays {\it (12)}. One dimensional
(1D) geometries are much more difficult to realize, due to the
unavoidable presence of on site frustration charges that substantially
modify the ideal phase diagram. 1D JJ arrays with neutral superfluids
(such as BEC), on the other hand, can be
accurately tailored, and open the possibility to observe directly
several remarkable phenomena not accessible to other systems {\it
  (14)}. First experiments with BECs held in a vertical optical
lattice have shown the spatial and temporal coherence of condensate
waves emitted at different heights of the gravitational field {\it
  (10)}. More recently, the degree of phase coherence among different
sites of the array {\it (15)} has been explored in the BEC ground
state configuration.

We report on the realization of a 1D array of JJs by
loading a BEC into an optical lattice potential generated by a
standing wave laser field.  The current-phase dynamics, driven by an
external harmonic oscillator potential provided by an external
magnetic field, maps on a pendulum-like equation and we have performed
a measurement of the critical Josephson current as a function of the
interwell potentials created by the light field.

The experimental apparatus has been described in detail elsewhere {\it
  (11)}.  We produce BECs of $^{87}$Rb atoms in the Zeeman state
$m_F=-1$ of the hyperfine level $F=1$ confined by a cylindrically
symmetric harmonic magnetic trap and a blue detuned laser standing
wave, superimposed on the axis of the magnetic trap.  In essence, the
cylindrical magnetic trap is divided into an array of disk shaped
traps by the light standing wave.  The axial and radial frequencies of
the magnetic trap are, $\omega_x=2\pi \times 9$\,Hz and $\omega_r = 2
\pi \times 92$\,Hz respectively.  By varying the intensity of the
superimposed laser beam (detuned 150 GHz to the blue of the D1
transition at $\lambda=795$\,nm) up to $14$\,mW/mm$^2$ we can vary the
interwell barrier energy $V_0$ from $0$ to $5~E_R$ where $E_R = h^2 /
2 m \lambda^2$ is the recoil energy of an atom (of mass $m$) absorbing
one of the lattice photons {\it (16)}.  The BEC is prepared by loading
$\sim 5 \times 10^8$ atoms in the magnetic trap and cooling the sample
through radio-frequency-forced evaporation until a substantial
fraction of condensed atoms is produced.  We then switch on the laser
standing-wave and continue the evaporation ramp until no thermal
component is experimentally visible. This ensures that the system
reaches the ground state of the combined trap.  The BEC splits in the
wells of the optical array: the distance between the wells is $\lambda
/ 2$ and $\sim 200$ wells are typically occupied, with $\sim 1000$
atoms in each well.  The interwell barrier energy $V_0$, and therefore
the tunneling rate, are controlled by varying the intensity of the
laser, which is chosen to be much higher than the condensate chemical
potential $\mu$.  $\mu$ ranges between $\mu \approx 0.10 \,
V_0$ for $V_0 = 2 \, E_R$ and  $\mu \approx 0.04 \, V_0$ for $V_0 = 5 \,
E_R$.  Each couple of condensates in neighbouring wells therefore
realizes a bosonic JJ, with a critical current $I_c$ depending on the
laser intensity.

In a more formal way we can decompose the condensate order parameter
that depends on position $\vec{r}$ and time $t$ 
as a sum of wave functions localized in each well of the periodic
potential (tight-binding approximation):
\begin{equation}
\label{ord-par}
\Psi(\vec{r},t)=\sqrt{N_T} \sum_j \psi_j(t) \Phi_j(\vec{r})
\end{equation}
where $N_T$ is the total number of atoms and $\psi_j=\sqrt{n_j(t)} \,
e^{i \phi_j (t)}$ is the $j$-th amplitude, with the fractional
population $n_j=N_j/N_T$ and the number of particles $N_j$ and the
phase $\phi_j$ in the trap $j$. This assumption relies on the fact
that the height of the interwell barriers is much higher than the
chemical potential. We will prove by a variational calculation that
this assumption is verified in most of the range of our experimental
parameters {\it (17)}.  The wave function $\Phi_j(\vec{r})$ of the
condensate in the $j$-th site of the array overlaps in the barrier
region with the wavefunctions $\Phi_{j \pm 1}$ of the condensates in
the neighbour sites. Therefore, the system realizes an array of weakly
coupled condensates, whose equation of motion satisfies a discrete
non-linear Schr\"odinger equation {\it (18)}:
\begin{equation}
\label{dnls}
 i  \hbar \frac{\partial \psi_n}{\partial t} = - K   
(\psi_{n-1}+\psi_{n+1}) + (\epsilon_n+ \Lambda \mid \psi_n \mid ^2)\psi_n  
\end{equation}
where $\epsilon_n= \Omega n^2$, $\Omega=\frac{1}{2} m \omega_x^2
\big(\frac{\lambda}{2} \big)^2 = 1.54 \times 10^{-5} E_R$, and $U =
g_0 N_T\int d\vec{r} \, \Phi_j^4$. The tunneling rate is proportional
to $K \simeq - \int d\vec{r} \, \big[ \frac{\hbar^2}{2m} \vec{\nabla}
\Phi_j \cdot \vec{\nabla} \Phi_{j+1} + \Phi_j V_{ext} \Phi_{j+1} \big]
$.  A simple variational estimate, assuming a gaussian profile for the
condensates in each trap, gives for $V_0 = 3 \, E_R$ the values $K
\sim 0.07 \, E_R$, $U \sim 12 \, E_R$ and a chemical potential $\mu
\sim 0.06 \, V_0$ that is much lower than the interwell potential
$V_0$.  We observe that the wavefunctions $\Phi_j$, as well as $K$,
depend on the height of the energy barrier $V_0$.

Eq. (\ref{dnls}) is a discrete non-linear 
Schr\"odinger equation (DNLSE) in a parabolic external potential,
conserving both 
the Hamiltonian 
${\cal{H}} = 
{\sum_j} [ - K ( \psi_j \psi^\ast_{j+1} + \psi^\ast_j
\psi_{j+1} )
+ \epsilon_j \mid
\psi_j\mid^2 + {U \over 2} \mid\psi_j\mid^4]$ and the norm 
$\sum_j n_j=1$. 

\begin{figure}
\begin{center}
\includegraphics[width=12cm]{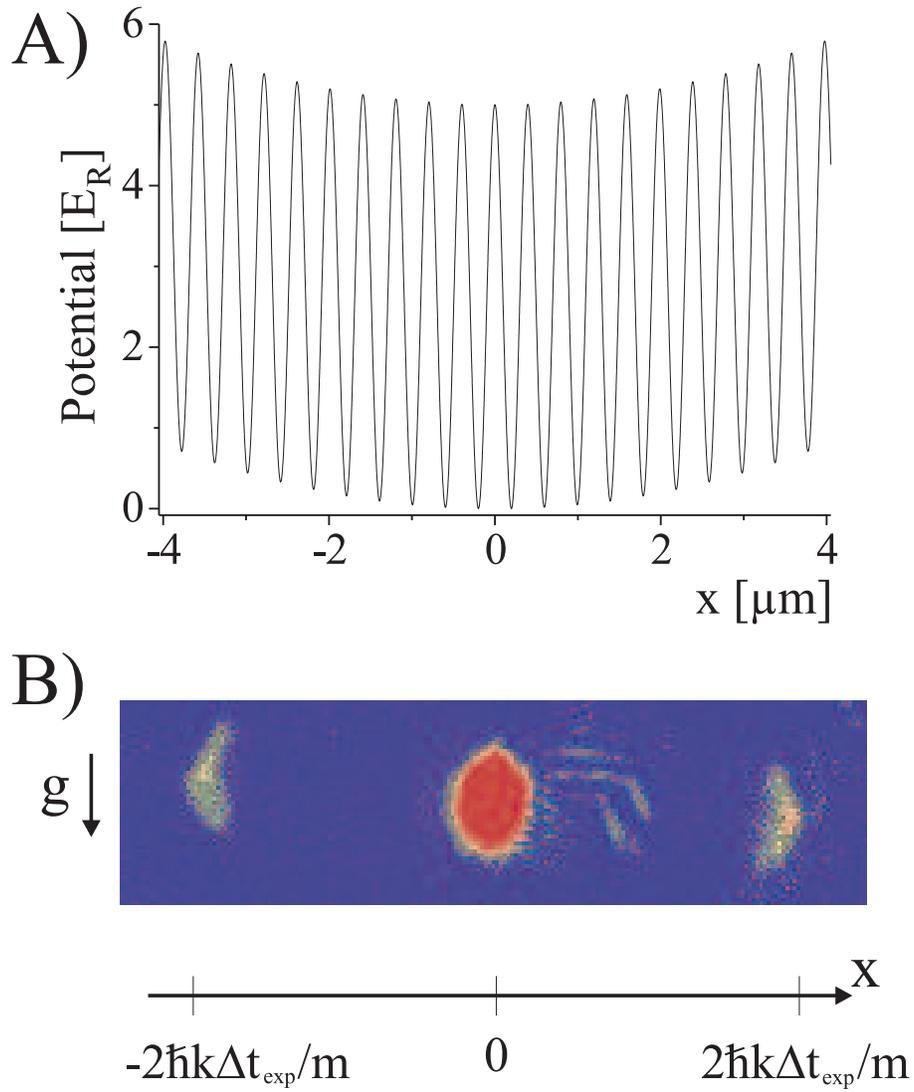}
\vspace{1cm}
\caption{A) Combined potential of the optical lattice and the magnetic
  trap in the axial direction. The curvature of the magnetic potential
  is exaggerated by a factor of 100 for clarity.  B) Absorption image
  of the BEC released from the combined trap. The expansion time was
  26.5\,ms and the optical potential height was $5 \,E_R$.  }
\label{trepic}
\end{center}   
\end{figure}

Although we can approximate the condensates in each lattice site as
having their own wave functions, tunneling between adjacent wells locks
all the different condensates in phase. As a result, when the
condensates are released from the combined trap, they will show an
interference pattern.  This pattern consists of a central peak plus a
symmetric comb of equally spaced peaks separated by $\pm 2 \hbar k_l
t_{exp} / m$ where $k_l$ is the wave vector of the trapping laser and
$t_{exp}$ is the expansion time.  In practice one can think of the far
field intensity distribution of a linear array of dipole antennas all
emitting with the same phase.  A complementary point of view is to
regard the density distribution after expansion as the Fourier
transform of the trapped one, i.e.\ the momentum distribution {\it
  (19)}.  It is easy to show that the sum of De Broglie waves
corresponding to momentum states integer multiples of $\pm 2 \hbar
k_l$ is the sum of localized wavefunctions of Eq.  (\ref{ord-par}).
The expanded cloud density distribution (Fig.\ref{trepic}) consists of
three distinct atomic clouds spaced by $\sim 306$\,$\mu$m$\simeq 2
\hbar k_l t_{exp} / m$ with the two external clouds corresponding to
the first order interference peaks, each containing roughly $10 \%$ of
the total number of atoms.  The interference pattern therefore
provides us with information about the relative phase of the different
condensates {\it (15,20)}; indeed, by repeating the experiment with
thermal clouds, even with a temperature considerably lower than the
interwell potential, we did not observe the interference pattern.

This situation is different from the Bragg diffraction
of a condensate released from a harmonic magnetic trap {\it (21)}
where the condensate is diffracted by a laser standing-wave. In our
case it is the ground state of the combined magnetic harmonic trap
plus optical periodic potential that by expansion produces an
interference pattern. For the time scales of our experiment the
relative intensities of the three interference peaks do not depend on
the time the atoms spend in the optical potential indicating that the
steady state system has been reached. In absence of external
perturbations the condensate remains in the state described by Eq.
(\ref{ord-par}) with a lifetime of $\sim 0.3$\,s at the maximum light
power, limited by scattering of light from the laser standing-wave.

In the ground state configuration the Bose-Einstein condensates are
distributed among the sites at the bottom of the parabolic trap. If we
suddenly displace the magnetic trap along the lattice axis by a small
distance $\sim 30$\,$\mu$m (the dimension of the array is
$\sim100$\,$\mu$m) the cloud will be out of equilibrium and will start
to move. As the potential energy that we give to the cloud is still
smaller than the interwell barrier each condensate can move along the
magnetic field only by tunneling through the barriers. A collective
motion can only be established at the price of an overall phase
coherence among the condensates. In other words, the relative phases
among all adjacent sites should remain locked together in order to
preserve the ordering of the collective motion. The locking of the
relative phases will again show up in the expanded cloud
interferogram.

\begin{figure}
\begin{center}
\includegraphics[width=12cm]{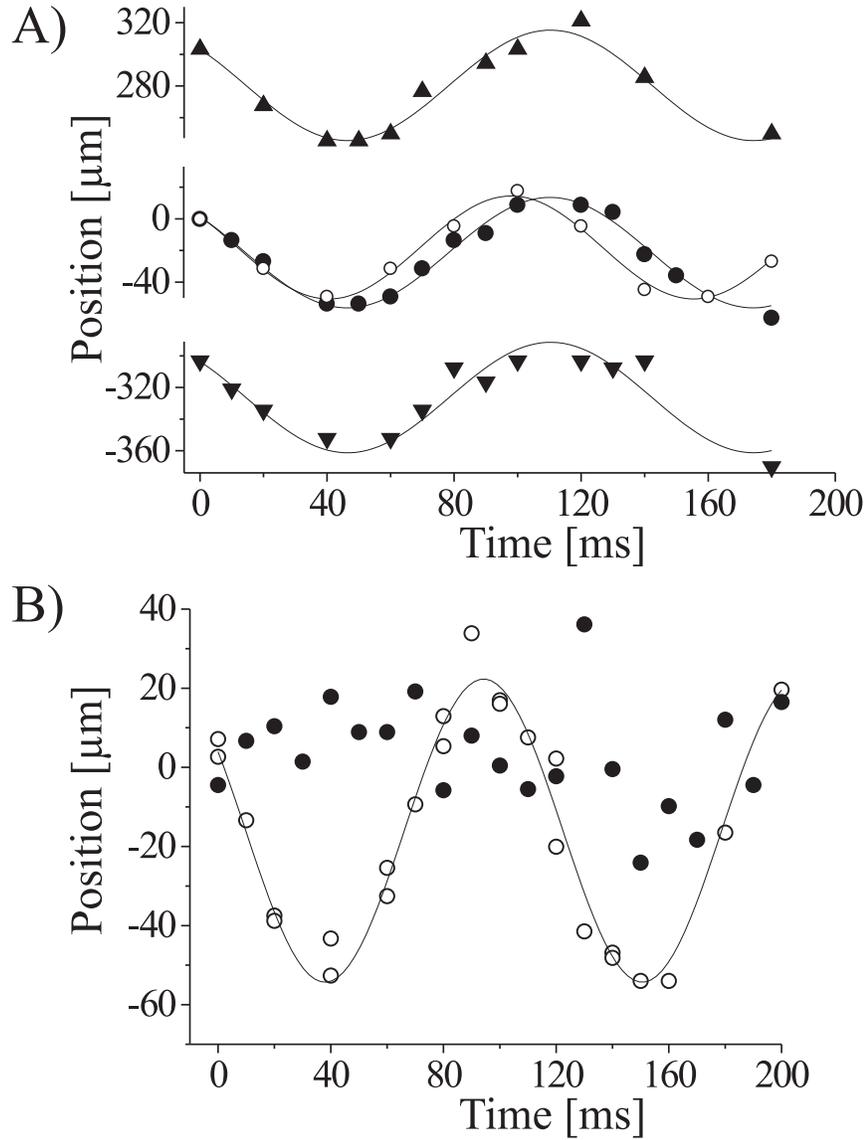}
\vspace{1cm}
\caption{A) Center of mass positions of the three peaks in the
  interferogram of the expanded condensate as a function of the time
  spent in the combined trap after displacement of the magnetic field.
  Up and down triangles correspond to the first order peaks, filled
  circles correspond to the central peak.  Open circles show the
  center of mass position of the BEC in absence of the optical
  lattice.  The continuous lines are the fits to the data.  B) Center
  of mass positions of the thermal cloud as a function of time spent
  in the displaced magnetic trap with the standing wave turned on
  (filled circles) and off (open circles).}
\label{oscilla}
\end{center}   
\end{figure}

For displacements that are not very large, we observe a coherent collective
oscillation of the condensates; i.e.\ we see the three peaks of the
interferogram of the expanded condensates oscillating in phase thus
showing that the quantum mechanical phase is maintained over the
entire condensate (Fig.\ref{oscilla}). In Fig. 2A,
we show the positions of the three peaks as a function of time spent
in the combined trap after the displacement of the magnetic trap,
compared with the motion of the condensate in the same displaced
magnetic trap but in absence of the optical standing wave (we refer to
this as ``harmonic'' oscillation).  The motion performed by the center
of mass of the condensate is an undamped oscillation at a
substantially lower frequency than in the ``harmonic'' case. We will
comment on this frequency shift later in the text; we would like now
to further stress the coherent nature of the oscillation. To do so we
repeat the same experiment with a thermal cloud. In this case,
although individual atoms are allowed to tunnel through the barriers,
no macroscopic phase is present in the cloud and no motion of the
center of mass should be observed.  The center of mass positions of
the thermal clouds are also reported in Fig.\ref{oscilla}B together
with the ``harmonic'' oscillation of the same cloud in absence of the
optical potential.  As can be seen, the thermal cloud does not
move from its original position in presence of the optical lattice.
Indeed, if a mixed cloud is used only the condensate fraction starts
to oscillate while the thermal component remains static, with the
interaction of the two eventually leading to a damping of the
condensate motion.

We now turn back to the discussion of the frequency reduction observed
in the oscillation of the pure condensate in presence of the optical
lattice. The current flowing through the junction between two quantum
fluids has a maximum value, the critical Josephson current $I_c$,
which is directly proportional to the tunneling rate.  The
existence of such a condition essentially limits the maximum velocity
at which the condensate can flow through the interwell barriers and
therefore reduces the frequency of the oscillations. As a consequence,
we expect a dependence of the oscillation frequency on the optical
potential through the tunneling rate.

To formalize the above reasoning, we rewrite the DNLSE (Eq.
\ref{dnls}) in terms of the canonically conjugate population/phase
variables, therefore enlightening its equivalence with the Josephson
equations for a one dimensional junction array:
\begin{equation}\label{n-phi} 
\left\{\begin{array}{ll} 
\hbar \dot{n}_j = 2K \sqrt{n_j n_{j-1}} \sin{(\phi_j-\phi_{j-1})} - 2K \sqrt{n_j n_{j+1}} \sin{(\phi_{j+1}-\phi_j)}                   \\
\hbar \dot{\phi}_j=-U n_j -\Omega j^2 + K \sqrt{n_{j-1} / n_j} \cos{(\phi_j-\phi_{j-1})} + 
K \sqrt{n_{j+1} / n_j} \cos{(\phi_{j+1}-\phi_j)}
\end{array}
\right.
\end{equation}

It is useful to introduce collective coordinates {\it (18)}: the
center of mass $\xi(t)$ and the dispersion $\sigma(t)$ are defined,
respectively, as $\xi(t)=\sum_j j n_j$ and $\sigma^2(t)=\sum_j j^2 n_j -
\xi^2$.  From Eq. \ref{n-phi}, we have $\hbar \dot{\xi} = 2K \sum_j
\sqrt{n_j n_{j+1}} \sin{(\phi_{j+1}-\phi_j)}$.  As the number of atoms
is large, the ``kinetic'' energy term of DNLSE is small respect to the
potential and nonlinear terms, and the population density profile is
simply given by an inverted discrete parabolic profile, centered
around $\xi$ {\it (22)}: $n_j(t)= \frac{\mu - \Omega
  (j-\xi)^2}{U}$; furthermore $\frac{d}{dt} \sigma^2=0$.

During the dynamical evolution, the relative phases across the
junctions $\phi_{j+1}-\phi_j \equiv \Delta \phi(t)$ remain locked
together to the same (oscillating) value.  This has been verified by
numerically studying the Fourier transform $\psi_k=\sum_j \psi_j e^{i
  k j}$; from the experimental point of view this means that the
expanded condensate continues to show the three peaks of the
interferogram of Fig.\ref{trepic}. Therefore in these collective
coordinate the current-phase relation is given by
\begin{equation}
\label{nc-deltaphi} 
\left\{\begin{array}{ll} 
\hbar \frac{d}{dt}\xi(t) = 2K ~ \sin{\Delta \phi(t)}   \\
\hbar \frac{d}{dt}\Delta \phi(t)= - ~ m \omega_x^2 
\big( \frac{\lambda}{2} \big)^2 ~ \xi(t)
\end{array}
\right.
\end{equation}
which, in analogy with the case of a superconducting Josephson
junction (in the resistively shunted junction model {\it (1,2)}) and
with the case of $^3He$ {\it (5)}, is a pendulum equation with the
relative phase $\Delta \phi$ corresponding to the angle to a vertical
axis and the center of mass $\xi$ being the corresponding angular
momentum.  The
current-phase dynamics does not depend explicitly on the interatomic
interaction.  This allows us to study regimes with a number of
condensate atoms spanning over different order of magnitude, which is
different from the configuration considered in {\it (10)} where
nonlinear effects would dephase the collective dynamics. However, it
is clear that the nonlinear interaction is crucial to determining the
superfluid nature of the coupled condensates, by locking the overall
phase coherence against perturbations.

From Eq. \ref{nc-deltaphi} we can see that the small amplitude
oscillation frequency $\omega_l$ of the current $I \equiv
N_T\frac{d}{dt}\xi$ gives a direct measurement of the critical Josephson
current $I_c \equiv 2 K N_T / \hbar$ and, therefore, of the atomic
tunneling rate of each condensate through the barriers.  The critical
current is related to the frequency $\omega$ of the atomic
oscillations in the lattice and to the frequency $\omega_x$ of the
condensate oscillations in absence of the periodic field by the
relation
\begin{equation}
\label{critical-current}
I_c = \frac{4 \hbar N_T}{m \lambda^2} \left( \frac{\omega}{\omega_x} \right)^2.
\end{equation}

\begin{figure}
\begin{center}
\includegraphics[width=12cm]{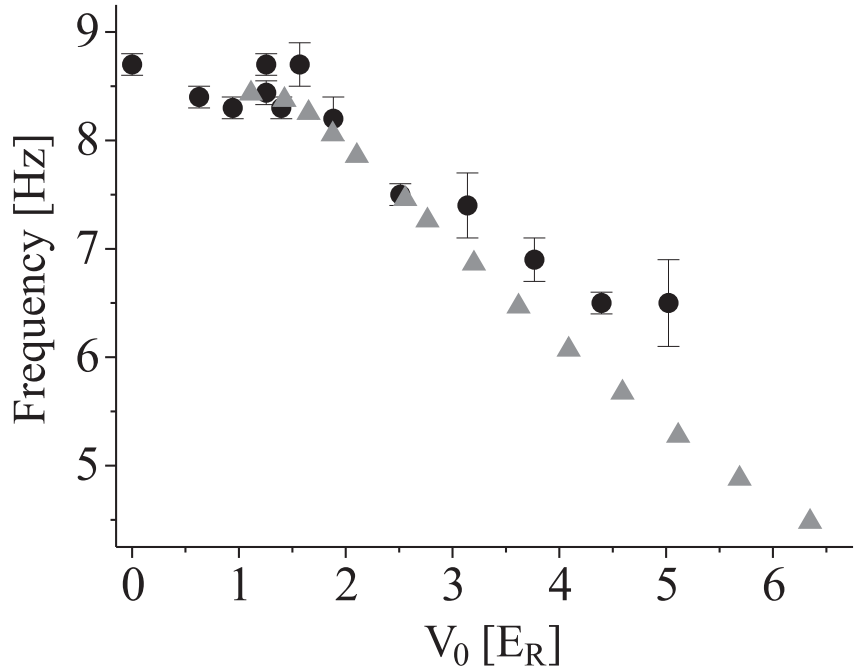}
\vspace{1cm}
\caption{The frequency of the atomic current in the array of Josephson
  junctions as a function of the interwell potential height.
  Experimental data (circles) are compared to the calculated values
  (triangles). Each experimenatl data point was taken after a complete
  oscillation in the displaced magnetic trap. The oscillation was then
  fitted with a sine function giving the corresponding frequency
  (error bars are the standard deviation of the data from the fit.}
\label{shift}
\end{center}   
\end{figure}

Figure \ref{shift} shows the experimental values of the oscillation
frequencies together with the result of a variational calculation
based on Eq. \ref{n-phi}. It must be noted that, due to mean field
interactions, in our system a bound state exists in the lattice only
for potentials higher than $\sim E_R$; frequency shifts for lower
potential heights are better explained in terms of the effective mass
$\frac{1}{m_{eff}}=\frac{\partial^2 {\cal {H}}}{\partial k^2}$ of the
system {\it (11)}.

Increasing the initial angular momentum $\xi_{0}$, the pendulum
librations become anharmonic, and can eventually reach the value
$\Delta \phi_{max}=\pi /2$. The system becomes dynamically unstable,
and the phase coherence is lost after a transient time (the
interference patterns washes out).  In this regime, the pendulum
analogy breaks down and a different dynamical picture would emerge.

With this work we have verified that the BEC's dynamics on a lattice
is governed by a discrete, non-linear, Schr\"odinger equation. This
equation is common to a large class of discrete non-linear systems,
including polarons, optical fibers, and biological molecules {\it
  (23)}, thus opening up interdisciplinary research. The phase
rigidity among different wells can be probed against thermal
fluctuations to test various theories of decoherence {\it (24)}. One
could study the role of collective dynamical
modes in the creation of solitons and kinks of the type described in
{\it (18,25,26)} (see also {\it (23)}) and more generally the routes to
quantum phase transitions in nonhomogeneous, low-dimensional systems.

\newpage
\noindent
1. A. Barone, in {\it Quantum Mesoscopic Phenomena and Mesoscopic
  Devices in Microelectronics} I. O. Kulik, R.  Ellialtioglu, Eds.
(Kluwer Academic Publishers 2000) pp.301-320;
\newline
2. A. Barone, G.  Paterno, {\it Physics and Applications of the
  Josephson Effect} (Wiley, New York 1982).
\newline
3. B. D. Josephson, {\it Phys.~Lett.} {\bf 1}, 251 (1962).
\newline
4. O. Avenel, E. Varoquaux, {\it Phys.~Rev.~Lett.} {\bf 60}, 416
(1988).
\newline
5. S. V. Pereverzev, S. Backaus, A. Loshak, J. C. Davis, R. E.
Packard, {\it Nature} {\bf 388}, 449 (1997).
\newline
6. {\it Bose-Einstein Condensation in Atomic Gases}, M.~Inguscio,
C.~E.~Wieman, S.~Stringari, Eds. (IOS Press Amsterdam, Oxford, Tokio,
Washington, 1999).
\newline
7. {\it Bose-Einstein Condensates and Atom Lasers}, S.~Martellucci,
A.~N.~Chester, A.~Aspect, M.~Inguscio, Eds.  (Kluwer Academic/Plenum
Publishers 2000).
\newline
8. F. Dalfovo, S. Giorgini, L. P. Pitaevskii, S. Stringari, {\it Rev.
  Mod. Phys.} {\bf 71}, 463 (1999).
\newline
9. A. J. Legget {\it Rev. Mod. Phys.} {\bf 73}, 307 (2001)
\newline
10. B. P. Anderson, M. A. Kasevich, {\it Science} {\bf 282}, 1686
(1998).
\newline
11. S. Burger {\it et al.}, {\it Phys.~Rev.~Lett.} {\bf 86}, 4447
(2001).
\newline
12. R. Fazio, H. van der Zant, cond-mat/0011152.
\newline
13. Y. Makhlin, G. Sch\"{o}n, A. Shnirman, {\it Rev.~Mod.~Phys.} {\bf
  73}, 357 (2001).
\newline
14. A. Smerzi, S. Fantoni, S. Giovannazzi, S. R. Shenoy, {\it
  Phys.~Rev.~Lett.} {\bf 79}, 4950 (1997).
\newline
15. C. Orzel, A. K. Tuchman, M. L. Fenselau, M. Yasuda, M. A.
Kasevich, {\it Science} {\bf 291}, 2386 (2001).
\newline
16. The value of the optical potential used in all the variational
calculations was calibrated by performing Bragg diffraction
experiments on the BEC released from the harmonic trap.  The
experimental result deviates from the potential calculated from the
measured laser power mainly because of alignment imperfections.
\newline
17. The validity of the tight-binding approximation is also based on
the fact that the tunneling of atoms in the higher energy band is
energetically forbidden. Because the gap is $\sim 3 E_R$, the potential
energy $\frac{1}{2} m \omega_x^2 \big( \frac{\lambda}{2} \big)^2 j^2$
for that would require $j \sim 500$, i.e.\ displacements three times
larger than the condensate dimensions.
\newline
18. A. Trombettoni, A. Smerzi, {\it Phys.~Rev.~Lett.} {\bf 86}, 2353
(2001).
\newline
19. The expanded density distribution reproduces the momentum
distribution for expansion times much longer than the inverse of the
trapping frequencies and if the nonlinear terms in the Schr\"odinger
equations (the mean field) can be neglected during the expansion. The
trapping frequencies of the single traps in the array are of the order
of a few kHz while the expansion time is 26.5\,ms, so the first
assumption is readily verified. The question of neglecting the mean
field in the first part of the expansion when the density is still
comparable to the original condensate is more delicate. However this
will only affect the shape of the single interference peaks and not
the overall interference pattern.
\newline
20. M. Greiner, I. Bloch, O. Mandel, T. W.  H\"{a}nsch, T. Esslinger,
cond-mat/0105105.
\newline
21. M. Kozuma {\it et al.}, {\it Phys.~Rev.~Lett.}  {\bf 82}, 871
(1999).
\newline
22. This is the discrete analoug of the ``Thomas-Fermi'' approximation
for the continuous Gross-Pitaevski equation with an external parabolic
potential. In this limit, as will be shown below, the dynamics does
not depend explicitely on the non-linear interatomic interaction,
which only governs the overall shape. Our collective mode, indeed, 
can be seen as the discrete analog of the dipole mode in the
continuous Gross-Pitaevskii equation, which frequency depends only on
the parameters of the external harmonic trap.
\newline
23. A. C. Scott {\it Nonlinear Science: Emergence and
    dynamics of coherent structures}, Oxford University Press (1999).
\newline
24. W. Zurek, {\it Phys. Today} {\bf 44}, 36 (1991).
\newline
25. A. Sanchez, A. R. Bishop, SIAM Rev. {\bf 40}, 579 (1988).
\newline
26. F. Kh. Abdullaev, B. B. Baizakov, S. A. Darmanyan, 
V. V. Konotop, and M. Salerno cond-mat/0106042
\newline
27. This work has been supported by the Cofinanziamento MURST, by
  the European Community under Contract No.  HPRI-CT-1999-00111 and
  HPRN-CT-2000-00125, and by the INFM Progetto di Ricerca Avanzata
  "Photonmatter".  We thank M.~Kasevich and S.~Stringari for helpful
  discussion.  A.\,S. and A.\,T. wish to thank the LENS for the kind
  hospitality during the realisation of this work.

\end{document}